\documentclass[journal,twoside,web]{ieeecolor}
\usepackage{tmi}
\usepackage{cite}
\usepackage{array}
\usepackage{tabularx}
\usepackage{amsmath,amssymb,amsfonts}
\usepackage{booktabs}
\usepackage{algorithmic}
\usepackage{multirow}
\usepackage{graphicx}
\usepackage{float}
\usepackage{xcolor}
\usepackage{textcomp}
\usepackage{placeins}
\usepackage{longtable}
\usepackage{url}
\definecolor{lightblue}{RGB}{173,216,230}
\definecolor{myblue}{RGB}{0,0,255}
\usepackage{hyperref}
\hypersetup{
	hypertex=true,
	colorlinks=true,
	linkcolor=myblue,
	anchorcolor=myblue,
	citecolor=myblue
}
\usepackage{caption} 

\def\BibTeX{{\rm B\kern-.05em{\sc i\kern-.025em b}\kern-.08em
		T\kern-.1667em\lower.7ex\hbox{E}\kern-.125emX}}
\begin{document}
	\title{
	     PWD: Prior-Guided and Wavelet-Enhanced Diffusion Model for Fast Limited-Angle imaging on Dental CT
	}
	\author{Yi Liu, Yiyang Wen, Zekun Zhou, Junqi Ma, Linghang Wang, Yucheng Yao, Liu Shi, \\ Qiegen Liu,  \IEEEmembership{Senior Member, IEEE}\vspace{-3 em}
		\thanks{This work was supported by the National Natural Science Foundation of China (621220033, 62201193). (Y. Liu and Y. Wen are co-first authors.) (Co-corresponding authors: L. Shi and Q. Liu.)}
		\thanks{Y. Liu, Y. Wen, L. Shi, and Q. Liu are with the School of Information Engineering, Z. Zhou is with School of Mathematics and Computer Sciences, Nanchang University, Nanchang 330031, China (email: \{shiliu, liuqiegen\}@ncu.edu.cn, \{liuyi, wenyiyang, zekunzhou\}@email.ncu.edu.cn).}
		\thanks{J. Ma, L. Wang, and Y. Yao are with YOFO Medical Technology Co., Ltd., Hefei 230088, China W(info@yofomedical.com).}
	}
	\maketitle
	\begin{abstract}
	Generative diffusion models have received increasing attention in medical imaging, particularly in limited-angle computed tomography (LACT). Standard diffusion models achieve high-quality image reconstruction but require a large number of sampling steps during inference, resulting in substantial computational overhead. Although skip-sampling strategies have been proposed to improve efficiency, they often lead to loss of fine structural details. To address this issue, we propose a prior information embedding and wavelet feature fusion fast sampling diffusion model for LACT reconstruction. The PWD enables efficient sampling while preserving reconstruction fidelity in LACT, and effectively mitigates the degradation typically introduced by skip-sampling. Specifically, during the training phase, PWD maps the distribution of LACT images to that of fully sampled target images, enabling the model to learn structural correspondences between them. During inference, the LACT image serves as an explicit prior to guide the sampling trajectory, allowing for high-quality reconstruction with significantly fewer steps. In addition, PWD performs multi-scale feature fusion in the wavelet domain, effectively enhancing the reconstruction of fine details by leveraging both low-frequency and high-frequency information. Quantitative and qualitative evaluations on clinical dental arch CBCT and periapical datasets demonstrate that PWD outperforms existing methods under the same sampling condition. Using only 50 sampling steps, PWD achieves at least 1.7 dB improvement in PSNR and 10\% gain in SSIM.
	
	\end{abstract}
	
	\begin{IEEEkeywords}
		Limited-angle CT reconstruction, guided diffusion, fast sampling, wavelet convolution.
	\end{IEEEkeywords}
	
	\section{Introduction}
	\label{sec:introduction}
	\IEEEPARstart{L}IMITED-ANGLE computed tomography (LACT) has become a significant focus in medical imaging research due to its rapid data acquisition and reduced radiation dose. In practical applications, system configuration constraints often result in insufficient angular coverage, a situation commonly encountered in fixed CT systems\cite{zhang2021directional}, mammography\cite{guo2023spectral2spectral}, short-exposure and motion-compensated acquisitions\cite{wurfl2018deep}, dental CT\cite{hu2022dior}, and C-arm CT\cite{bachar2007carm}. However, LACT suffers from incomplete projection data, which often leads to severe artifacts, blurred details, and structural distortions in the reconstructed images. This makes LACT reconstruction one of the critical challenges in medical imaging.
	
	In iterative reconstruction approaches, compressed sensing techniques and total variation regularization were widely adopted \cite{TV,hu2017improved,luo2017image}. However, in the context of LACT reconstruction, these methods often led to structural blurring and residual artifacts. Deep learning-based approaches achieved notable progress in several studies \cite{xie2022limited, germer2023limited, danielyan2011bm3d, elad2006image, pan2024iterative}, offering fast inference as a key advantage. Nevertheless, they remained vulnerable to missing projection data, and their reconstruction performance degraded under noise interference or varying scan angles.

	\begin{figure}[t]
		\centering	
 		\includegraphics[width=0.5\textwidth]{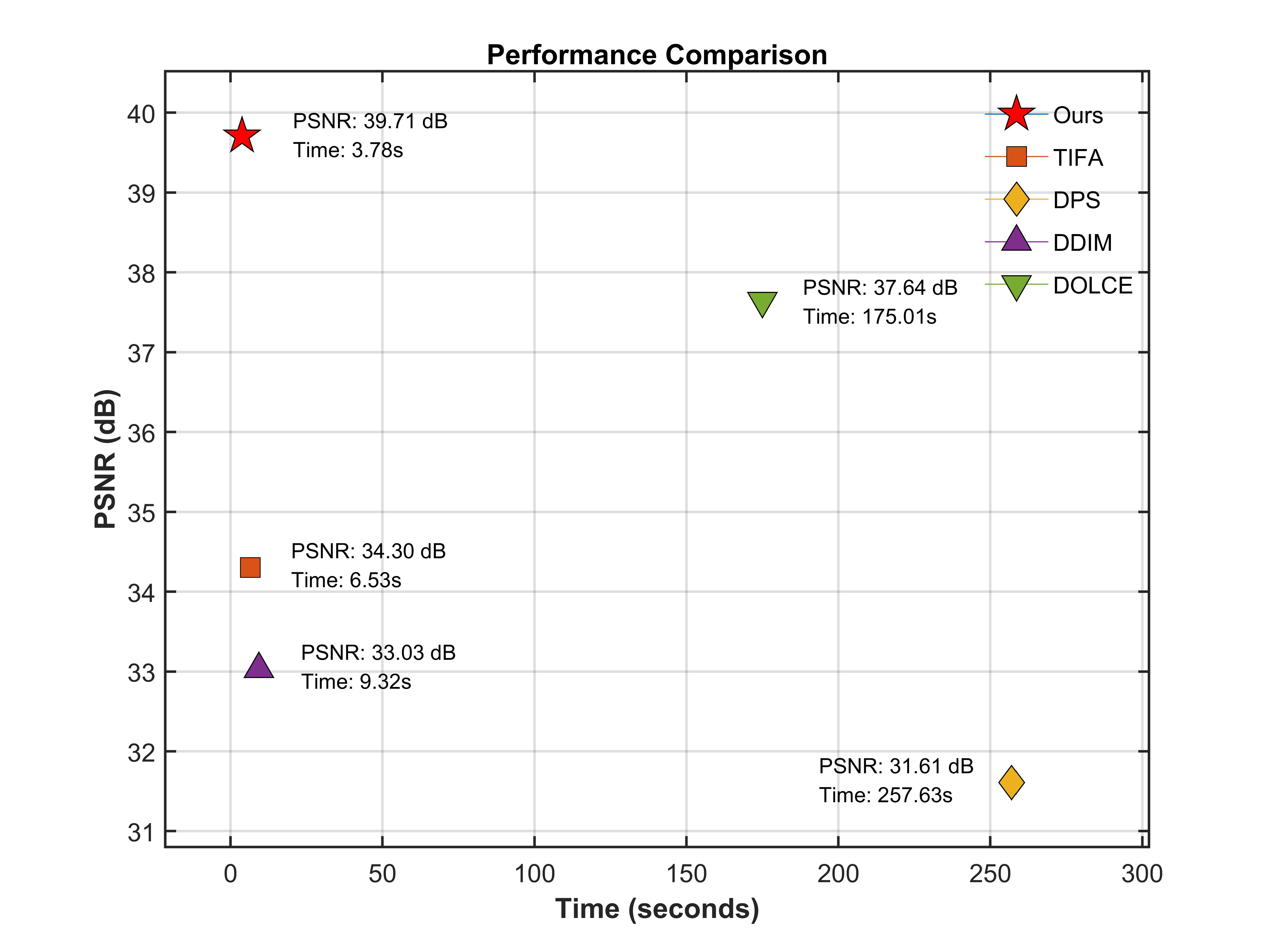} 
		\caption{Quantitative comparison of PWD and other diffusion model methods (TIFA\cite{wang2024time}, DPS\cite{chung2022diffusion}, DOLCE\cite{liu2023dolce}, DDIM\cite{song2020denoising}) in LACT reconstruction quality and efficiency, with scanning angle of 90°.}
		\label{time_psnr}
		\vspace{-15pt}
	\end{figure}
		
	In recent years, generative diffusion models achieved remarkable progress in image reconstruction tasks \cite{cao2024survey, shen2025efficient, zeng2025diffusion, xiao2021tackling, masip2023continual}. Within the field of medical image reconstruction, two primary directions drove their advancement. One line of research focused on integrating diffusion models with traditional reconstruction algorithms, leveraging physical modeling and optimization strategies to enhance prior information and improve reconstruction quality \cite{lu2024pridediff,du2024dper,wang2020deep}. The other line emphasized improving sampling efficiency and model generalization, such as by introducing frequency-domain representations, simplifying network structures, or incorporating Transformer-based architectures to accelerate inference \cite{xie2024prior,peebles2023scalable,2018limited}. Against this background, diffusion models emerged as a promising solution for LACT reconstruction owing to their probabilistic formulation, which facilitates the integration of prior knowledge into the reconstruction process. For instance, Liu \textit{et al.}~\cite{liu2023dolce} proposed DOLCE, which jointly optimized data fidelity and generative priors under a probabilistic diffusion framework, substantially reducing streak artifacts. Wang \textit{et al.}~\cite{wang2024time} developed TIFA, which applied timestep inversion and resampling to accelerate sampling and improve reconstruction performance in LACT. Zhang \textit{et al.}~\cite{zhang2024wavelet} introduced WISM, which combined wavelet-domain decomposition with a multi-channel diffusion model to further enhance reconstruction quality. Although these methods demonstrated improvements in image quality, standard diffusion models remained computationally intensive due to their reliance on numerous sampling steps during inference. To address this limitation, recent studies explored frequency-domain modeling \cite{yang2023diffusion}\cite{jiang2025frequency} and the design of efficient samplers \cite{jiang2024fast}\cite{jiang2025fast}. However, two critical challenges persisted. First, the structural priors inherent in LACT images were not fully exploited. Second, aggressive sampling strategies often resulted in the loss of fine details and the persistence of artifacts. Achieving efficient sampling while preserving high reconstruction quality remained a fundamental challenge in this domain.

	To improve the aforementioned challenges, this study proposes PWD, an accelerated sampling strategy based on prior information embedding and wavelet-domain feature fusion, aiming at enhancing the reconstruction performance of LACT. Specifically, during the training phase, PWD incorporates LACT prior information by mapping the distribution of LACT images to that of the corresponding target images, enabling the model to effectively capture spatial structural features under limited-angle constraints. Additionally, a multi-scale wavelet-domain feature fusion mechanism is introduced to fully exploit both low-frequency structural details and high-frequency edge information, thereby enhancing the model’s reconstruction capability. During inference, a guided skip-sampling strategy is adopted, wherein the LACT prior image serves as structural guidance to mitigate potential detail loss and artifact persistence associated with aggressive sampling reduction, thus ensuring stable and high-quality reconstruction. \textcolor{subsectioncolor}{Fig.}~\ref{time_psnr} presents a performance comparison between PWD and existing diffusion-based methods in terms of reconstruction accuracy and computational efficiency. The main contributions of this work are summarized as follows:

	\indent $\bullet$ \emph{\textbf{Prior Information Embedding:}} A prior-guided strategy is introduced to map the distribution of LACT images to that of the target images. This design enables the model to capture the structural correspondence between the input and the target, thereby improving the stability and accuracy of reconstruction.
	
	\indent $\bullet$ \emph{\textbf{Wavelet Feature Enhancement:}} A wavelet feature fusion framework is constructed to incorporate structural information from multiple frequency bands into the data modeling process. This design significantly improves the reconstruction of high-frequency details such as edges and textures, enabling high-fidelity image recovery.
	
	\indent $\bullet$ \emph{\textbf{Guided Fast Sampling:}} To improve the issue of detail degradation and artifact persistence commonly observed in conventional skip-sampling schemes, a guided fast sampling mechanism is introduced. By explicitly incorporating LACT prior information into the sampling process, the proposed method effectively constrains the sampling trajectory, leading to convergence towards higher quality solutions.

	The structure of this paper is organized as follows. Section II reviews related work on LACT reconstruction. Section III presents the motivation and overall workflow of the proposed PWD. Section IV presents the experimental results, followed by further discussions and conclusions in Section V and VI.

	
	\section{Preliminary}
	{
		\subsection{LACT Reconstruction}
		
		In computed tomography, the forward projection model is generally expressed as:
		\begin{equation}
			\mathbf{y} = \mathbf{A}x + \boldsymbol{\zeta},
			\label{eq:ct_forward_model}
		\end{equation}
		where $x \in \mathbb{R}^n$ denotes the target image, $\mathbf{A} \in \mathbb{R}^{d \times n}$ is the system matrix corresponding to the full-angle Radon transform, and $\boldsymbol{\zeta}$ denotes the measurement noise. The projection vector $\mathbf{y} \in \mathbb{R}^d$ is acquired over a complete angular range.
		
		In the LACT setting, only a subset of the full-angle projections is available. This incomplete measurement process can be modeled as:
		\begin{equation}
			\mathbf{y} = \mathbf{M} \odot (\mathbf{A} x) + \boldsymbol{\zeta},
			\label{eq:lact_forward_model}
		\end{equation}
		where $\mathbf{M} \in \{0, 1\}^d$ is a binary angular sampling mask that indicates the available projection views, and $\odot$ denotes element-wise multiplication. The condition $|\mathbf{M}| \ll d$ typically holds, reflecting the severely truncated angular coverage in LACT, and resulting in an intrinsically incomplete observation $\mathbf{y}$. To recover $x$ from such undersampled data, the reconstruction problem is formulated as the following regularized inverse problem:
		\begin{equation}
			\hat{x} = \arg\min_{x} \left\| \mathbf{M} \odot (\mathbf{A} x) - \mathbf{y} \right\|_2^2 + \lambda \mathcal{R}(x),
			\label{eq:lact_inverse_problem}
		\end{equation}
		where $\mathcal{R}(x)$ denotes a regularization term. The inverse problem is inherently ill-posed due to the non-trivial null space of the masked system operator $\mathbf{M} \odot \mathbf{A}$, which lacks full rank under limited-angle acquisition. This rank deficiency leads to non-uniqueness and instability in the solution. Moreover, the missing angular projections hinder the recovery of directional structures. As a result, traditional analytic reconstruction methods, such as filtered back-projection (FBP), often suffer from severe streak artifacts and structural distortions in the reconstructed images.
		
		To address the ill-posedness of LACT reconstruction, various strategies have been proposed. Traditional iterative algorithms, such as the FISTA\cite{beck2009fast}, enhance reconstruction stability by solving a regularized inverse problem. Deep learning-based methods, such as FBPConvNet\cite{jin2017deep}, optimize FBP outputs through convolutional neural network post-processing. More recently, generative approaches have attracted attention. For instance, SPGAN\cite{yang2018low} performs end-to-end optimization via generative adversarial networks. Diffusion model-based methods such as DPS~\cite{chung2022diffusion} aim to reconstruct high-fidelity images directly from degraded projections via probabilistic generative modeling.
	}
	{
		\subsection{Diffusion Models}
		Diffusion models have recently emerged as a powerful class of generative models in the field of deep learning. Their core mechanism involves progressively adding noise to the data during a forward process, such that the data distribution gradually approaches a standard Gaussian. A neural network is then trained to reverse this process through iterative denoising, enabling the recovery of the original data distribution from pure noise. Representative examples include denoising diffusion probabilistic models (DDPMs)\cite{ho2020denoising} and score-based generative models based on stochastic differential equations (SDEs)\cite{song2020score}, both of which are unsupervised approaches that learn the underlying data distribution to generate high-quality samples.
			
		Despite their promising performance, a major limitation of diffusion models lies in their computational inefficiency. The sampling process typically requires hundreds to thousands of iterative steps to reconstruct a clean image from random noise, which significantly hampers practical deployment. To address this issue, several acceleration strategies have been proposed. Notably, the denoising diffusion implicit model (DDIM) \cite{song2020denoising} adopts a deterministic sampling mechanism to skip a large number of intermediate steps while preserving image fidelity. Model distillation techniques~\cite{meng2023distillation}\cite{salimans2022progressive} further reduce the sampling burden by extracting lightweight student models from pre-trained diffusion models, enabling an exponential reduction in the number of inference steps.
		
		Existing diffusion-based methods rely primarily on the statistical properties learned during training and treat them as the sole guidance for the generative process. In the actual reconstruction phase, these methods often lack effective integration of observed data or known prior information, thereby limiting their generalization ability and reconstruction quality in ill-posed scenarios. In the context of LACT reconstruction, incorporating known structural priors plays a critical role in enhancing reconstruction quality.
			
	}
	
	\section{Method}

	
	\subsection{Motivation}
	
\begin{figure}[htp]
	\centering
	\includegraphics[width=0.5\textwidth]{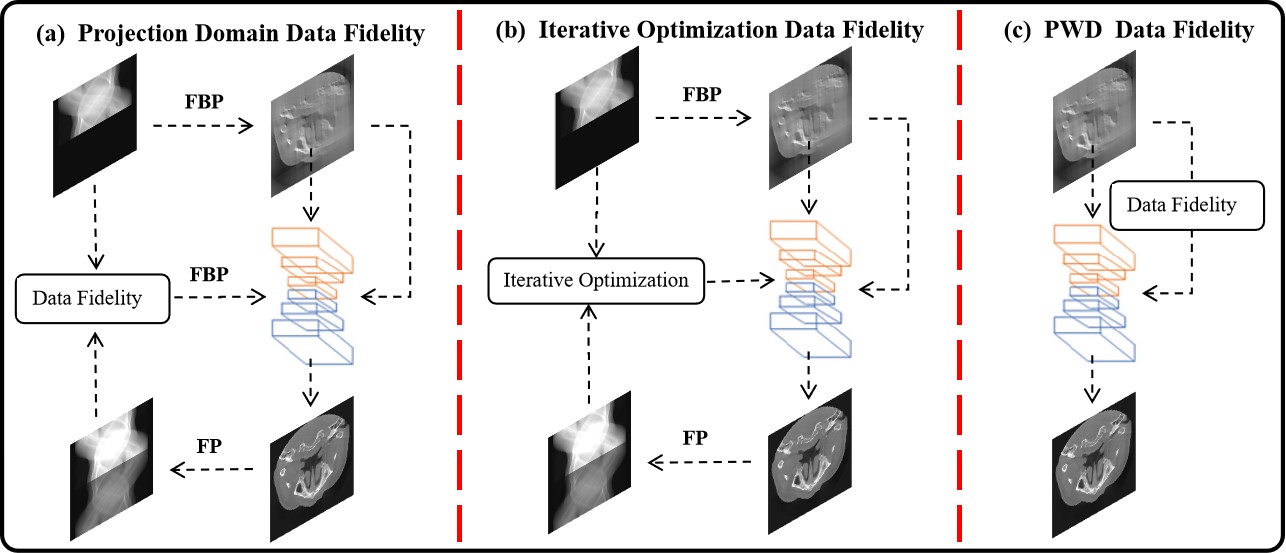} 
	
	\caption{The comparison of different data fidelity strategies. (a) shows projection-domain replacement-based fidelity, (b) depicts iterative algorithm-based data fidelity, (c) illustrates the data fidelity approach adopted by PWD.}
	\label{all}
	\vspace{-15pt}
	\end{figure}

	{
	In previous studies on LACT, image fidelity was typically maintained using two main categories of methods, as illustrated in Fig.~\textcolor{subsectioncolor}{\ref{all}} (a) and Fig.~\textcolor{subsectioncolor}{\ref{all}} (b). The first category enhances data consistency by integrating traditional reconstruction algorithms. For example, DOLCE~\cite{liu2023dolce} employs the accelerated proximal gradient method (APGM) for data consistency enforcement,  DPS~\cite{chung2022diffusion} combines deep image priors with regularization to improve reconstruction robustness. The second category explicitly completes missing projections in the projection domain. TIFA~\cite{wang2024time} performs projection completion by replacing parts of the current reconstruction with known projection data during each sampling iteration. These approaches often rely on full forward or backward projection operations, resulting in high computational cost.
		
	Due to the characteristics of limited-angle acquisition, the FBP reconstruction from the combined projections of available and missing angular ranges retains structural consistency with the reconstruction from full-angle projections. This property stems from the linearity of the FBP algorithm~\cite{beckmann2020error}. Based on this observation, as illustrated in Fig.~\textcolor{subsectioncolor}{\ref{all}} (c), the LACT image is embedded as fidelity information into the reconstruction process. In terms of computational complexity, this operation has a time complexity of $\mathcal{O}(1)$, which is significantly lower than that of conventional data consistency methods. The latter typically require full forward or backward projection operations, resulting in a computational complexity of at least $\mathcal{O}(n)$ and incurring higher computational costs. From the data constraints, introducing a structural prior image as a conditional constraint helps restrict the solution space during reconstruction, thereby improving the stability and accuracy of the results.
	}

	\begin{figure*}[htp]
		\centering
		\includegraphics[width=0.8\textwidth]{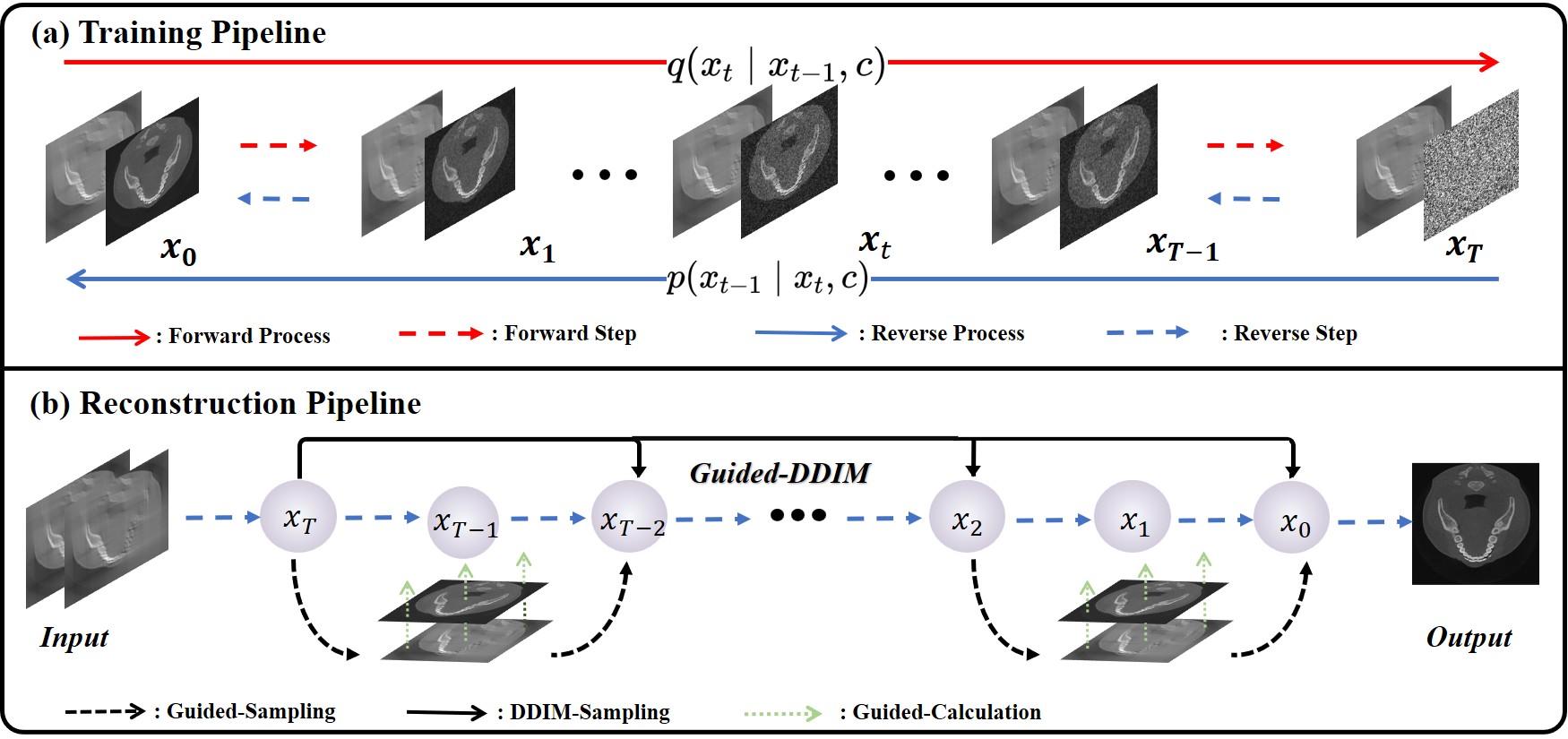} 
		
		\caption{Illustration of the training and sampling pipeline of PWD. (a) shows the training phase of PWD, (b) depicts the Guided-DDIM sampling strategy used during inference.}
		\label{forward and sample process}
		\vspace{-10pt}
	\end{figure*}
	

\subsection{Optimizing of PWD}

{
	\:\:\noindent\textbf{\textit{Prior Information Embedding: }}This section details the training strategy adopted in PWD. To enhance generative fidelity and generalization, PWD incorporates structural prior information \( c \) from LACT images into the diffusion model training, guiding the model to learn the structural correspondence between LACT inputs and fully sampled CT targets.
	
	Given a clean ground-truth image \( x_0 \in \mathbb{R}^{H \times W} \), Gaussian noise is added during training to produce the perturbed input \( x_t \), while the prior image \( c \in \mathbb{R}^{H \times W} \) remains unchanged. The forward diffusion process is defined as:
	
	\begin{equation}
		x_t = \sqrt{\bar{\alpha}_t} \, x_0 + \sqrt{1 - \bar{\alpha}_t} \, {\epsilon}_t, \quad {\epsilon}_t \sim \mathcal{N}(0, \mathbf{I}),
		\label{eq:forward_process}
	\end{equation}
	where \( \bar{\alpha}_t \) denotes the cumulative noise attenuation coefficient at timestep \( t \), and \( \epsilon \) is a standard Gaussian noise term with zero mean and unit variance. The denoising network is trained to approximate the added noise using a conditional model \( \epsilon_{\theta}(x_t, c, t) \), where \( x_t \) is the noisy input at timestep \( t \), and \( c \) provides structural guidance. By conditioning on the LACT prior \( c \), the model implicitly learns a mapping from the distribution of limited-angle projections to the distribution of fully sampled CT images. According to Bayes theorem, the gradient of the conditional log-likelihood can be decomposed as:
	\begin{equation}
		\nabla_{x_t} \log p(c \mid x_t) = \nabla_{x_t} \log p(x_t \mid c) - \nabla_{x_t} \log p(x_t),
		\label{eq:score_decomp}
	\end{equation}
	where \( p(c \mid x_t) \) denotes the likelihood of observing the prior given \( x_t \), while \( p(x_t \mid c) \) and \( p(x_t) \) represent the conditional and unconditional distributions of \( x_t \), respectively.
	
	In diffusion models, the score function is closely related to noise prediction. The unconditional score can be approximated as:
	\begin{equation}
		\nabla_{x_t} \log p(x_t) \approx -\frac{1}{1 - \bar{\alpha}_t} \, \epsilon_{\theta}(x_t),
	\end{equation}
	similarly, the conditional score is approximated by:
	\begin{equation}
		\nabla_{x_t} \log p(x_t \mid c) \approx -\frac{1}{1 - \bar{\alpha}_t} \, \epsilon_{\theta}(x_t, c),
	\end{equation}
	where \( \tilde{\epsilon}_{\theta}(x_t, c) \) denotes the noise estimated by the model under conditional guidance.
	Essentially, guiding noise prediction is equivalent to updating the sampling along the aforementioned gradient direction.
	
	By substituting these into Eq.~\eqref{eq:score_decomp}, we derive the relationship between conditional and unconditional predictions:
	\begin{equation}
		\epsilon_{\theta}(x_t, c) = \epsilon_{\theta}(x_t) - \sqrt{1 - \bar{\alpha}_t} \, \nabla_{x_t} \log p(c \mid x_t),
		\label{eq:score_guided_eps}
	\end{equation}
	the training objective minimizes the discrepancy between the predicted and true noise via the mean squared error:
	\begin{equation}
		\mathcal{L}(\theta) = \mathbb{E}_{x_0, c, t, \epsilon} \left[ \left\| \epsilon_{\theta}(x_t, c) - \epsilon_t \right\|_2^2 \right],
		\label{eq:loss}
	\end{equation}
	where \( \mathcal{L}(\theta) \) denotes the loss function, \( \epsilon_{\theta}(x_t, c, t) \) is the model’s prediction, and \( \epsilon_t \) is the true noise added at timestep \( t \).
	
	Finally, during inference, the learned conditional denoising model approximates the target distribution by adjusting the predicted noise:
	\begin{equation}
		\epsilon_{\theta} \propto \epsilon_{\theta}(x_t, c, t) + \sigma_t \nabla_{x_t} \log p(c \mid x_t),
		\label{eq:final_eps}
	\end{equation}
	this formulation allows the model to effectively leverage prior information throughout the reverse process, enabling structurally faithful and high-quality reconstructions in LACT tasks.
}

\begin{figure}[htb]
	\centering
	\includegraphics[width=0.5\textwidth]{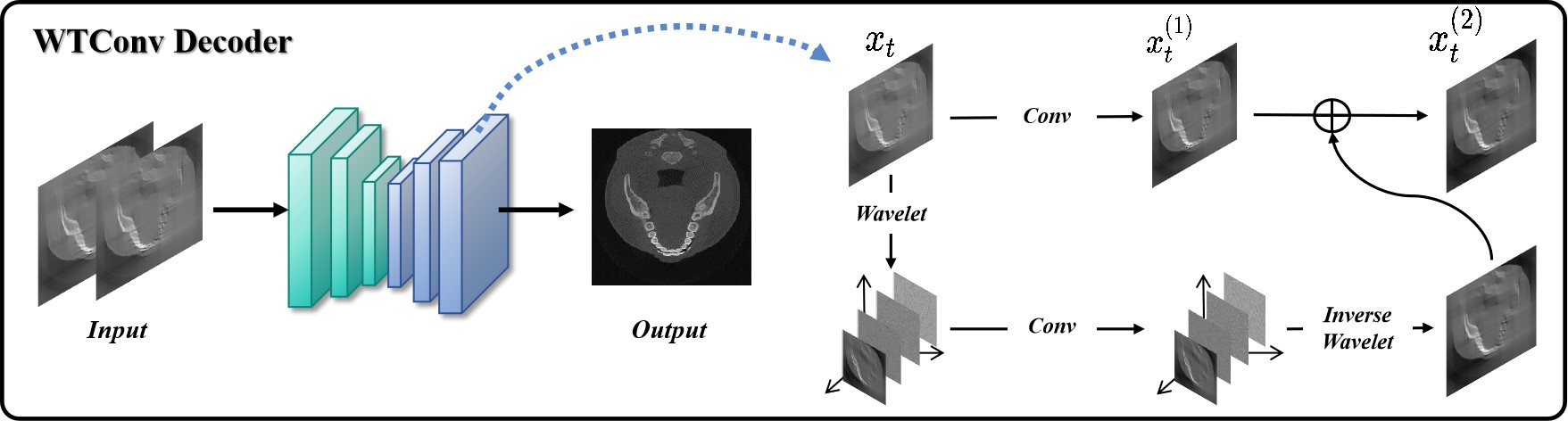} 		
	\caption{The processing procedure of WTConv. WTConv enables the model to capture both global structures and high-frequency details during the decoding process, theeby enhancing the quality of image.}
	\label{WTConv}
	\vspace{-10pt}
\end{figure}

 {
 	\:\:\noindent\textbf{\textit{WTConv Module: }}To improve the recovery of both global structures and fine details, we incorporate Wavelet Transform Convolution (WTConv) into the U-Net backbone of the diffusion model. The overall workflow is illustrated in Fig.~\textcolor{subsectioncolor}{\ref{WTConv}}.First, the input noisy image \( x_t \) is passed through a convolutional branch to extract coarse structural features:
 	\begin{equation}
 		x_t^{(1)} = {Conv}(x_t),
 	\end{equation}
 	
 	Simultaneously, \( x_t \) is decomposed into multi-scale frequency components using discrete wavelet transform (DWT):
 	\begin{equation}
 		\mathrm{WT}(x_t) = \left( x_t^{LL}, x_t^{LH}, x_t^{HL}, x_t^{HH} \right),
 	\end{equation}
 	where \( x_t^{LL} \) captures low-frequency structural information, while \( x_t^{LH} \), \( x_t^{HL} \), and \( x_t^{HH} \) encode high-frequency details in horizontal, vertical, and diagonal directions, respectively.
 	
 	Each wavelet subband is independently processed via depthwise separable convolution:
 	\begin{equation}
 		y_t^{(i)} = {DWConv}_{3 \times 3}(x_t^{(i)}), \quad i \in \{LL, LH, HL, HH\},
 	\end{equation}
 	where \( {DWConv}_{3 \times 3}(\cdot) \) denotes the depthwise separable convolution operation. The enhanced frequency features are fused using inverse wavelet transform:
 	\begin{equation}
 		\tilde{x}_t = \mathrm{iWT} \left( y_t^{LL},\; y_t^{LH},\; y_t^{HL},\; y_t^{HH} \right),
 	\end{equation}
 	
 	Finally, the outputs from the convolutional and wavelet branches are aggregated:
 	\begin{equation}
 		x_t^{(2)} =  {Conv}(x_t^{(1)} + \tilde{x}_t),
 	\end{equation}
 	where \( x_t^{(2)} \) denotes the final fused feature used for denoising prediction. This WTConv-enhanced design allows the model to simultaneously capture global structure and high-frequency details, significantly improving image reconstruction quality under limited-angle conditions.
    
 		
 	
 }
\subsection{Guided Fast Sampling}
{
	LACT reconstruction poses an ill-posed inverse problem due to severe projection truncation, as described in Eq.~\eqref{eq:lact_inverse_problem}. The null space of the masked operator $\mathbf{M} \odot \mathbf{A}$ lacks full rank, which compromises the stability and fidelity of the reconstructed image $x$. This motivates the integration of prior structural constraints during inference to regularize the solution space.
	
	To this end, we propose a guided fast sampling strategy that combines deterministic DDIM sampling with prior-based guidance. The overall pipeline is illustrated in Fig.~\textcolor{subsectioncolor}{\ref{forward and sample process} (b)}. Standard denoising diffusion implicit models (DDIM) update the latent variable $x_t$ as follows:
	\begin{equation}
		x_{t-1} = \underbrace{\frac{\alpha_{t-1}}{\alpha_t} \left( x_t - \sqrt{1 - \alpha_t} \, \epsilon_{\theta}(x_t)\right)}_{x_0^*} + \sqrt{1 - \alpha_{t-1}} \, \epsilon_{\theta}(x_t),
		\label{eq:ddim_update_refined}
	\end{equation}
	where $\epsilon_{\theta}(x_t)$ is the predicted noise and $\alpha_t$ is the variance coefficient. The first term $x_0^*$ represents the predicted clean image, while the second term adds calibrated noise for stochasticity. 
	
	However, in LACT settings, $x_0^*$ may lack reliable structural details due to incomplete measurements. To incorporate prior information, we define the posterior distribution as:
	\begin{equation}
		p(x_0 \mid x_t, c) \propto p(x_t \mid x_0) \cdot p(x_0 \mid c),
		\label{eq:bayesian_formulation}
	\end{equation}
	where $p(x_t \mid x_0)$ is determined by the diffusion model and $p(x_0 \mid c)$ is a Gaussian prior:
	\begin{equation}
		p(x_0 \mid c) = \mathcal{N}(x_0; c, \sigma_c^2 \mathbf{I}).
	\end{equation}
	Maximizing the log-posterior yields the guided estimate:
	\begin{equation}
		x_g = x_0^* + w \cdot (c - x_0^*),
		\label{eq:xg_refined}
	\end{equation}
	where $w = \sigma_0^2 / \sigma_c^2$ is a scalar weight. Substituting $x_g$ into Eq.~\eqref{eq:ddim_update_refined} gives the guided sampling:
	\begin{equation}
		x_{t-1} = \frac{\alpha_{t-1}}{\sqrt{\alpha_t}} \left[ x_0^* + w \cdot (c - x_0^*) \right] + \sqrt{1 - \alpha_{t-1}} \cdot \epsilon_{\theta}(x_t).
		\label{eq:final_guided_ddim}
	\end{equation}
	This can also be interpreted as minimizing a quadratic energy:
	\begin{equation}
		E(x_0) = \| x_0 - x_0^* \|_2^2 + w \cdot \| x_0 - c \|_2^2,
		\label{eq:refined_energy}
	\end{equation}
	whose closed-form solution corresponds to Eq.~\eqref{eq:xg_refined}.
	
	The guided term $c$ is derived from a low-cost LACT reconstruction (e.g., FBP), which empirically preserves coarse directional structure. As shown in Fig.~\textcolor{subsectioncolor}{\ref{all}}, summing limited- and missing-angle projections (or images) yields a reconstruction structurally consistent with full-angle results. This intrinsic property supports the use of $c$ as a structural prior.
}
 The workflow of PWD can be represented as Algorithm 1.
	
\vspace{0.5 em}
\noindent 
\begin{tabular}{p{0.47\textwidth}}
	\toprule
	
	\textbf{Algorithm 1 PWD} \\
	\midrule
	\textbf{Training Stage} \\
	\midrule
	\textbf{Dataset}: A paired dataset consisting of full-angle data and LACT data. \\
	
	1: \textbf{Repeat}
	
	2:\quad Sample a data pair $(x, c)$ from the dataset \\
	3: \quad $\epsilon_{\theta}(x_{t}, t)=\epsilon_{\theta}(x_{t}, c, t),$
	
	4:	\quad	$\mathcal{L}(\theta) = \mathbb{E}_{x_0, c, t, \epsilon} \left[ \left\| \epsilon_{\theta}(x_{t}, t) - \epsilon_t \right\|_2^2 \right]$
	
	5:\quad Take a gradient descent step on \\ 
	\quad \quad $\nabla_{\theta} \mathcal{L}(\theta) = \nabla_{\theta} \mathbb{E}_{x_0, c, t, \epsilon} \left[ \left\| \epsilon_{\theta}(x_{t}, t) - \epsilon_t \right\|_2^2 \right]$ \\
	
	6: \quad $\theta \leftarrow \theta - \eta \cdot \nabla_{\theta} \mathcal{L}(\theta)$ \\

	7: \textbf{Until Converged} \\
	
	8: \textbf{Trained PWD} \\
	\midrule
	\textbf{Reconstruction Stage} \\
	\midrule

	1: \textbf{Input:} $x_{t}$ , $\epsilon_{\theta}$ (Initial noise)\\
	
	2: \textbf{Output:} Reconstructed LACT : $x_0$ \\
	
	3: \textbf{Sample:} $x_{t} \sim \mathcal{N}(0, \mathbf{I})$ \\
	
	4: \textbf{For} $t = T$ to $0$ \textbf{do}: \\
	
	5: \quad $\epsilon_{\theta} \sim \mathcal{N}(0, \mathbf{I})$ \\
	
	6: \quad $x_{t-1} = \frac{\alpha_{t-1}}{\sqrt{\alpha_t}} \left[ x_0^* + w \cdot (c - x_0^*) \right] +\sqrt{1 - \alpha_{t-1}} \cdot \
	\epsilon_{\theta}(x_t)$ \\
	
	
	7: \textbf{End For} \\
	
	8: \textbf{Return:} $x_0$ \\
	
	\bottomrule
\end{tabular}

\vspace{1em}

	
	\section{Experiments}
	
	All reconstruction experiments in this study were implemented using the PyTorch and conducted on a workstation equipped with a single NVIDIA RTX 4090 GPU with 24 GB of memory. The Adam optimizer was employed with an initial learning rate set to $1\times10^{-4}$, and training was performed to minimize the Eq.~(\ref{eq:loss}). To comprehensively evaluate the performance of the proposed method, we compared it with several representative baseline approaches, including the (FBP\cite{kak2001principles}), supervised based methods such as FBPConvNet\cite{jin2017deep} and IRON\cite{pan2024iterative}, as well as diffusion based models including DPS\cite{chung2022diffusion}, DDIM\cite{song2020denoising}, DOLCE\cite{liu2023dolce}, and TIFA\cite{wang2024time}. All baseline methods were configured according to the parameter settings reported in their original publications, and we optimized their implementations as reasonably as possible to ensure a fair comparison. For quantitative evaluation, two widely adopted image quality metrics peak signal to noise ratio (PSNR) and structural similarity index (SSIM) were used to assess the fidelity and structural consistency of the reconstructed images.

	\subsection{Datasets Preparation}
\begin{figure}[htb]
		\centering
		\includegraphics[width=0.5\textwidth]{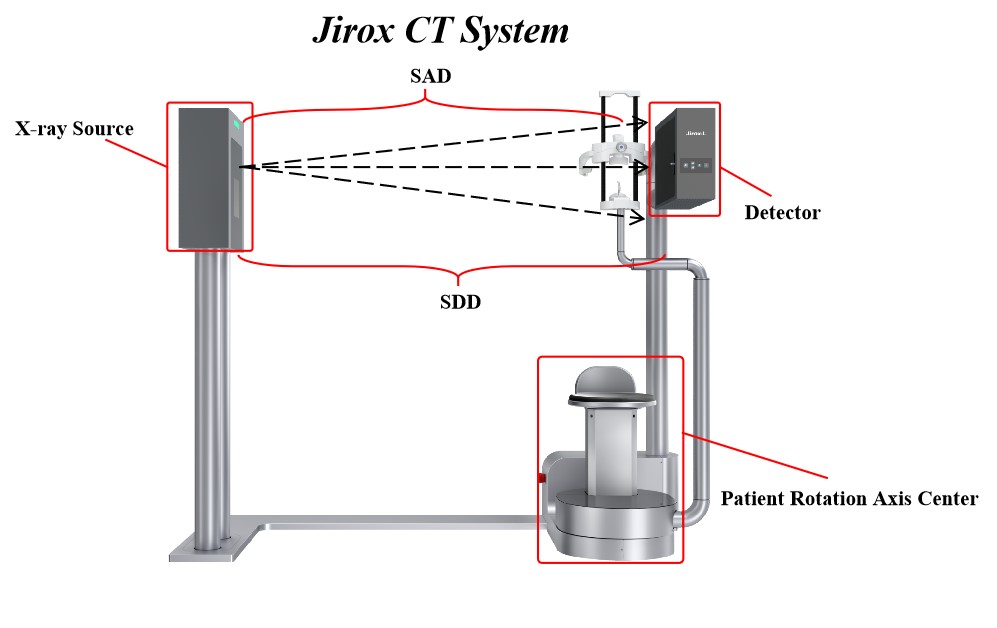} 
		
		\caption{The Jirox CT system, designed and manufactured by YOFO Medical Technology Co., Ltd., with its scanning geometry indicated.}
		\label{Jirox}
		\vspace{-5pt}
	\end{figure}

	All the data used in this study were acquired using the Jirox CT system, whose structural design is illustrated in \textcolor{subsectioncolor}{Fig.} \ref{Jirox}. The system was designed and manufactured by YOFO Medical Technology Co., Ltd. Unlike conventional dental CBCT systems, the Jirox CT adopts a source-to-detector distance (SDD) of 1700 mm and a source-to-axis distance (SAD) of 1500 mm. This configuration brings the cone-beam geometry closer to parallel-beam conditions, thereby improving reconstruction accuracy and image quality while enabling a larger field of view. Due to the extended SAD, the system is designed to allow patient self-rotation to complete the scanning process. The Jirox CT operates with an X-ray tube voltage of 100kV and a tube current of 6 mA. However, the large SID leads to significant X-ray attenuation during propagation, especially under fixed dose settings, resulting in low-dose imaging conditions and degraded signal quality. To mitigate this issue, an in-house low-dose reconstruction algorithm was employed to perform correction and recovery, thereby enhancing the final image quality. Considering the potential motion artifacts caused by patient self-rotation and its impact on practical usability, this study investigates the feasibility of limited-angle scanning based on the Jirox CT system.

	\subsubsection{Dental Arch Dataset}
	
	{
		This dataset consisted of 200 cases, each comprising 200 slices located at the dental arch. In total, the dataset included 40,000 full-view sinograms and their corresponding reconstruction results. The sinograms were of size 1536 $\times$ 1600, corresponding to 1600 uniformly sampled projections over the angular range of [0°, 360°]. The reconstructed images were of size 640 $\times$ 640. Following the [0°, 90°] and [0°, 120°] LACT simulation protocols, the dataset was repurposed to generate paired LACT samples from the full-view sinograms. Specifically, each data pair included a full-angle CT slice and the corresponding LACT slice. The dataset was partitioned into a training set of 180 cases and a test set of 20 cases, resulting in 36,000 training samples and 4,000 test samples.
		
	}

	\subsubsection{Periapical Dataset}
	
	{
		Considering the specific characteristics of dental Cone Beam CT (CBCT) and the requirements of practical system design, a dataset was constructed to reflect realistic scanning conditions. Specifically, 90° angular range data were retained from real CBCT acquisitions, and the remaining projections were discarded. Full-view CT images were reconstructed using FBP, from which paired periapical samples were extracted, as illustrated in \textcolor{subsectioncolor}{Fig.} \ref{Periapical}. This dataset more accurately reflected the actual scanning procedures of CBCT systems. It comprised 200 cases, each providing 400 axial slices. For each pair, one full-angle CT slice and its corresponding LACT reconstruction were included. The dataset was divided into a training set of 180 cases and a test set of 20 cases, yielding 72,000 training samples and 8,000 test samples.
		\begin{figure}[h]
			\centering
			\includegraphics[width=0.45\textwidth]{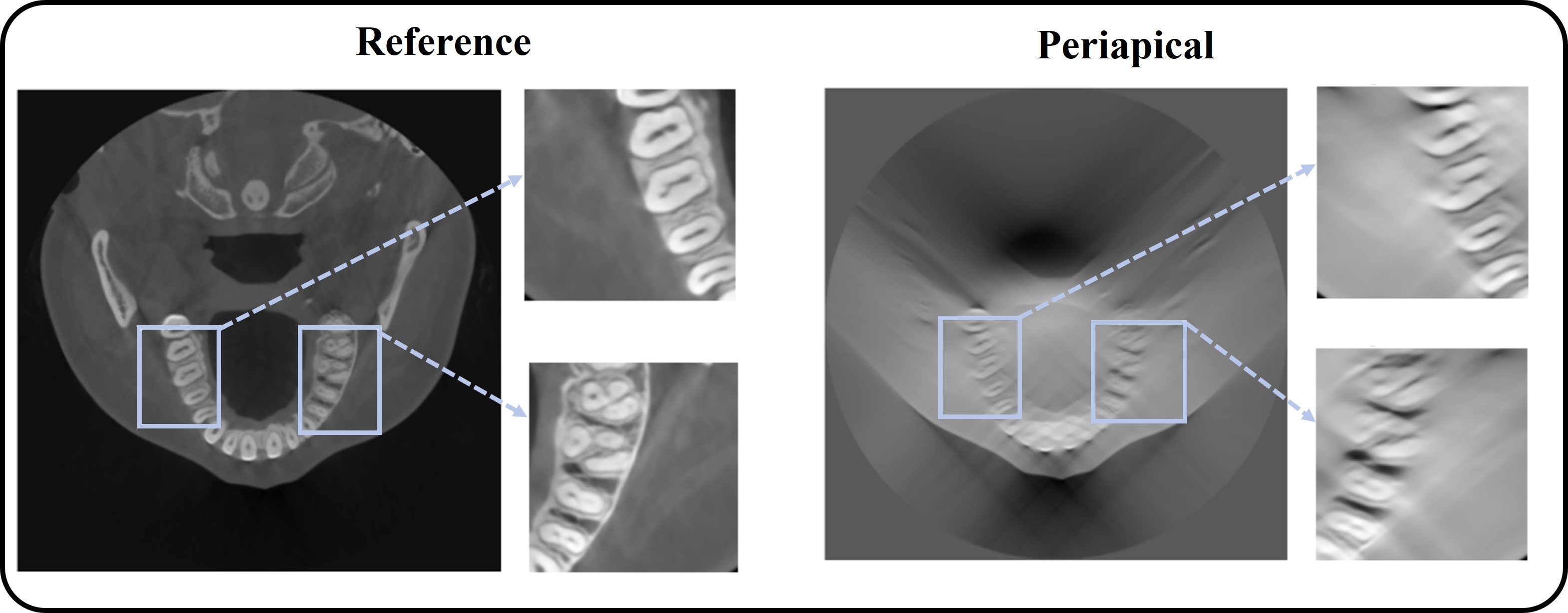} 
			\caption{Acquisition of periapical data under limited scanning conditions.}
			\label{Periapical}
			\vspace{-10pt}
		\end{figure}
		
	}
		
	\subsection{Dental Arch Reconstruction Study}
	{
		
		{
		LACT reconstruction is performed under two angular configurations of [0°, 90°] and [0°, 120°]. DPS and DOLCE adopt 1000 sampling steps. TIFA follows its original setting with 200 sampling steps. In contrast, DDIM and the proposed PWD use only 50 sampling steps. \textcolor{subsectioncolor}{Fig.} \ref{arch_90_case2} and \textcolor{subsectioncolor}{Fig.} \ref{arch_120_case1} present reconstruction results under both configurations. Under the 90° configuration, supervised methods such as FBPConvNet and IRON exhibit evident geometric distortions in structurally complex regions, failing to preserve anatomical fidelity. Diffusion-based approaches, including DPS, DDIM, DOLCE, and TIFA, yield more natural overall appearances but still suffer from blurred or incomplete edge reconstruction. In comparison, PWD maintains robustness under severe artifact conditions and accurately restores fine structural details. With a 120° scanning range, the increased projection coverage improves the overall performance of all diffusion-based methods. However, supervised approaches still show noticeable residual artifacts, especially in dental regions. Other generative methods generate globally consistent reconstructions but remain limited in recovering high-frequency details within the ROI. In contrast, PWD achieves higher accuracy and demonstrates superior structural fidelity and robustness across both angular settings. \textcolor{subsectioncolor}{Table}~\ref{arch_90_120_table1} summarizes the quantitative results for 10 randomly selected slices under [0°, 90°] and [0°, 120°] configurations. PWD improves PSNR by at least 1.74 dB and SSIM by at least 10\% under all test conditions. 
	
		}
		
		\begin{figure*}[htb]
			\centering
			\includegraphics[width=0.9\textwidth]{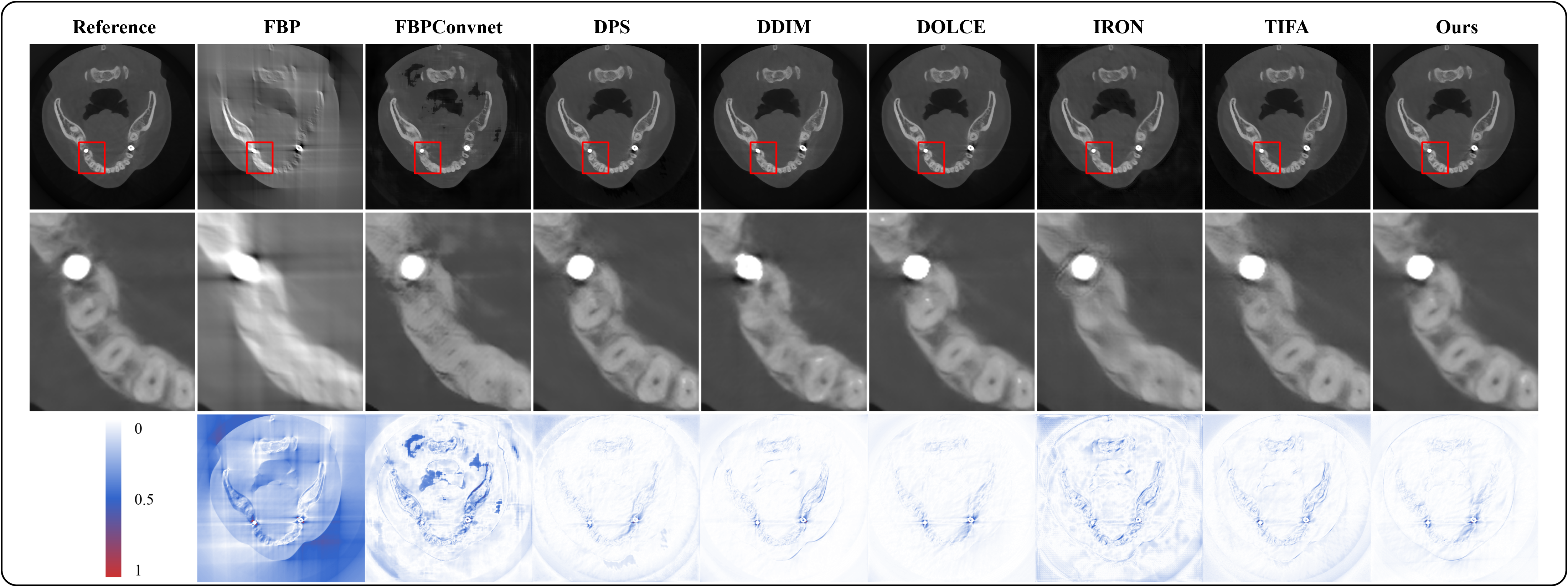} 
			
			\caption{Reconstruction results of different methods on the dental arch dataset with a scanning angle of 90°. The titles indicate the respective reconstruction methods. The second row displays the ROI regions of clinical interest, and the third row shows the difference maps between the reconstructed results and the reference images. The display window is [-900, 3700] HU.}
			\label{arch_90_case2}
			\vspace{-5pt}
		\end{figure*}
		
		\begin{figure*}[htb]
			\centering
			\includegraphics[width=0.9\textwidth]{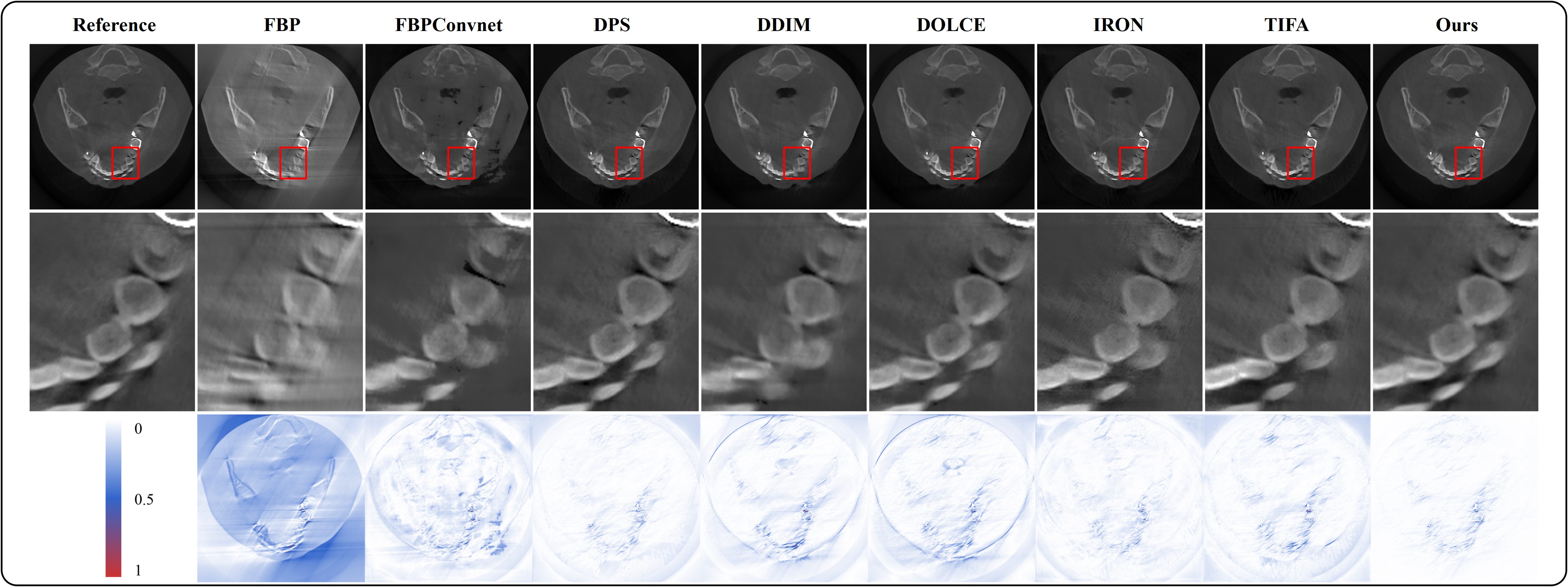} 
			
			\caption{Reconstruction results on the dental arch dataset at 120° scanning angle using different methods. The titles indicate the respective reconstruction methods. The second row shows the ROI regions of clinical interest. The third row presents the difference maps between the reconstructed images and the reference images. The display window is [-900, 3700] HU.}
			\label{arch_120_case1}
			\vspace{-5pt}
		\end{figure*}

		\begin{table}[htb]
			\centering
			\caption{Quantitative Evaluation of the PWD and Other Competition Methods on LACT 90° and 120° Dental Arch Datasets.}
			\label{arch_90_120_table1}
			\begin{tabular}{c|cc|cc}
				\toprule
				\multirow{2}{*}{Method} & \multicolumn{2}{c|}{$[0^\circ, 90^\circ]$} & \multicolumn{2}{c}{$[0^\circ, 120^\circ]$} \\
				\cmidrule(lr){2-3} \cmidrule(lr){4-5}
				& PSNR & SSIM & PSNR & SSIM \\
				\midrule
				FBP & 22.56 & 0.3639 & 25.63 & 0.5001 \\
				FBPConvnet & 28.50 & 0.8778 & 29.63 & 0.8946 \\
				DPS   & 31.61 & 0.6072 & 35.01 & 0.7211 \\
				DDIM & 33.03 & 0.8185 & 34.78 & 0.8372 \\
				IRON & 33.82 & 0.9291 & 36.37 & 0.9486\\
				TIFA & 37.30 & 0.8268 & 39.14 & 0.9392\\
				DOLCE  & 37.64 & 0.8878 & 40.11 & 0.9468 \\
				\textbf{Ours}    & \textbf{39.71} & \textbf{0.9646} & \textbf{41.85} & \textbf{0.9672} \\
				\bottomrule
			\end{tabular}
		\vspace{-10pt}
		\end{table}


	}
	
	\subsection{Periapical Reconstruction Study}
	
\begin{figure*}[htp]
		\centering
		\includegraphics[width=0.9\textwidth]{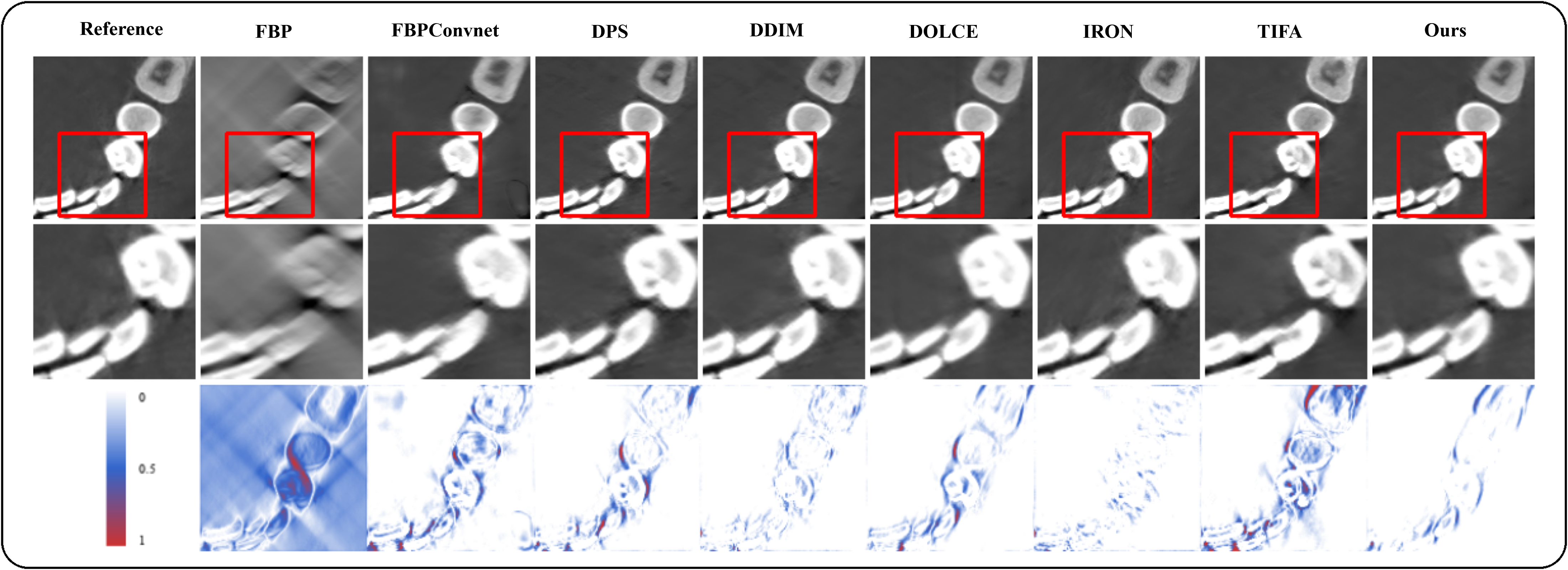} 
		\caption{ Reconstruction results of different methods on the periapical dataset with a scanning angle of 90°. The titles indicate the respective reconstruction methods. The second row shows magnified details of dental structures. The third row displays residual maps comparing the reconstruction results to the Reference. The display window is [-900, 3500] HU.}
		\label{veneer_90}
		\vspace{-5pt}
	\end{figure*}

\begin{figure*}[htp]
		\centering	
		\includegraphics[width=0.9\textwidth]{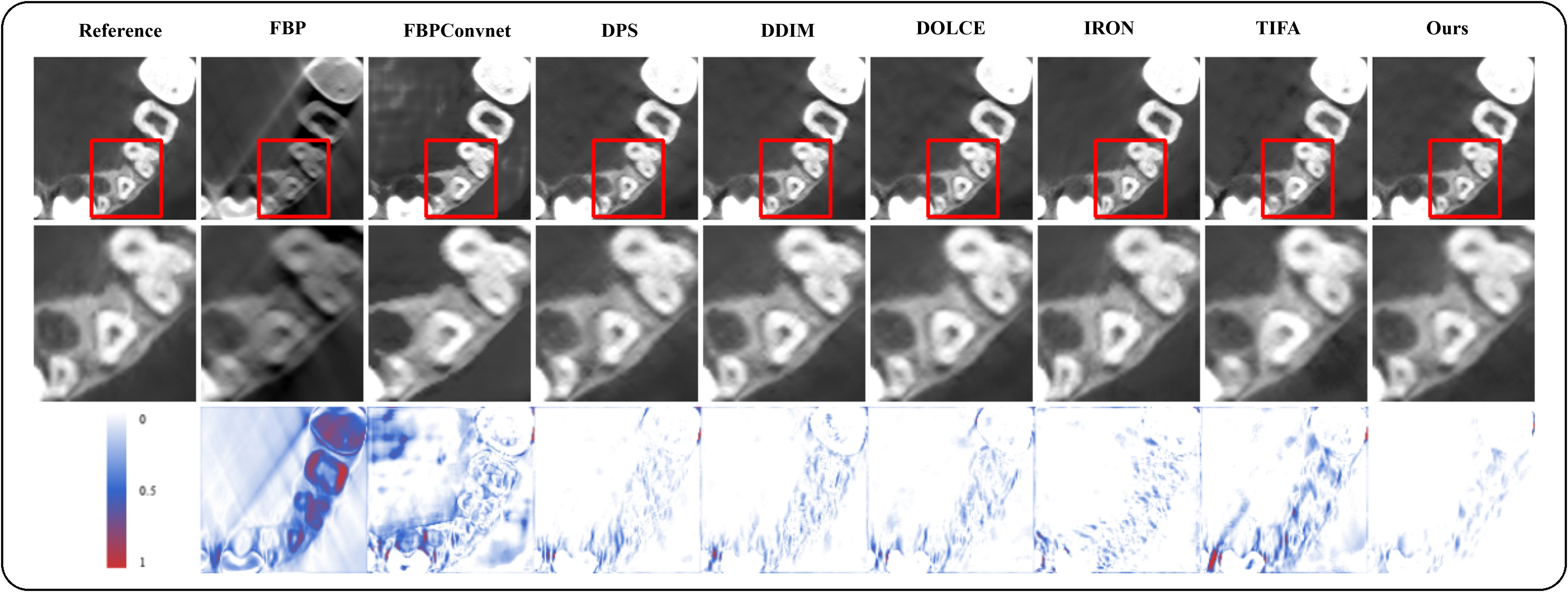} 
		
		\caption{ Reconstruction results of different methods on the periapical dataset with a scanning angle of 120°. The titles indicate the respective reconstruction methods. The second row shows magnified details of dental structures. The third row displays residual maps comparing the reconstruction results to the Reference. The display window is [-900, 3500] HU.}
		\label{veneer_120}
		\vspace{-5pt}
	\end{figure*}

	{
		As illustrated in \textcolor{subsectioncolor}{Fig.} \ref{veneer_90} and \textcolor{subsectioncolor}{Fig.} \ref{veneer_120}, a comparative analysis is conducted on periapical dental data under scanning angles of [0°, 90°] and [0°, 120°]. Common distortions in this type of image include blurred tooth structures and contour deformation. Under the 90° configuration, the high X-ray absorption of teeth often leads to pronounced artifacts and deformation at the edges. Reconstruction results within the ROI show that PWD achieves superior recovery of high-frequency details compared to other methods. Under the 120° configuration, although all methods yield generally satisfactory reconstructions, residual maps reveal that PWD outperforms others in detail preservation and error suppression. The reconstructed images from PWD present edge structures and intensity distributions that more closely match the reference. \textcolor{subsectioncolor}{Table}~\ref{veneer 90 120 table} summarizes the quantitative metrics under both angular configurations, where PWD consistently achieves higher average PSNR and SSIM than competing approaches.
	}
	\begin{table}[h]
		\centering
		\caption{Quantitative Evaluation of the PWD and Other Competition Methods on LACT 90° and 120° Periapical Datasets.}
		\label{veneer 90 120 table}
		\begin{tabular}{c|cc|cc}
			\toprule
			\multirow{2}{*}{Method} & \multicolumn{2}{c|}{$[0^\circ, 90^\circ]$} & \multicolumn{2}{c}{$[0^\circ, 120^\circ]$} \\
			\cmidrule(lr){2-3} \cmidrule(lr){4-5}
			& PSNR & SSIM & PSNR & SSIM \\
			\midrule
			FBP & 15.25 & 0.3752 & 20.86 & 0.4676 \\
			FBPConvnet & 22.35 & 0.9004 & 22.93 & 0.9175 \\
			DPS   & 27.01 & 0.8979 & 32.67 & 0.9681 \\
			DDIM & 32.48 & 0.9540 & 34.89 & 0.9751 \\
			IRON & 37.46 & 0.9818 & 37.07 & 0.9833 \\
			TIFA & 38.90 & 0.9497 & 39.82 & 0.9679\\
			DOLCE  & 40.09 & 0.9841 & 41.76 & 0.9824 \\
			\textbf{Ours}    & \textbf{41.34} & \textbf{0.9870} & \textbf{42.09} & \textbf{0.9971} \\
			\bottomrule
		\end{tabular}
		\vspace{-10pt}
	\end{table}
	

	\subsection{Accelerated Sampling Study}
	
\begin{figure*}[htp]
		\vspace{-5pt}
		\centering
		\includegraphics[width=0.80\textwidth]{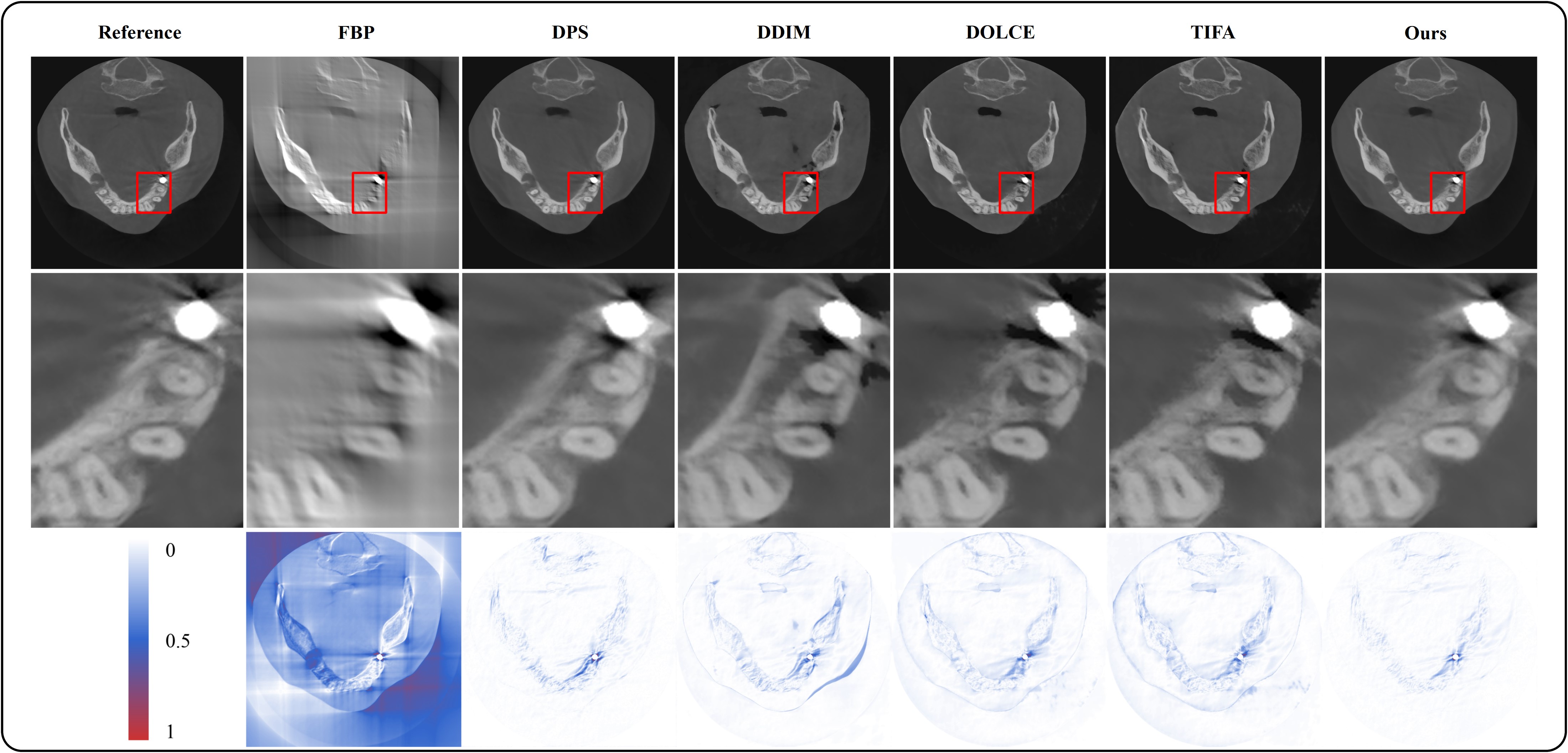} 
		\caption{Comparison of the effectiveness of PWD using the Guided-DDIM sampling approach and other diffusion modeling approaches sampling the DDIM sampling approach. The display window is set as [-400, 3000] HU.}
		\label{omparison of the effectiveness of PWD}
		\vspace{-10pt}
	\end{figure*}

	To further assess the effectiveness of the proposed method under accelerated sampling conditions, several representative diffusion models, including DPS, DDIM, and TIFA, were selected as comparative baselines. All methods were evaluated under an identical sampling setting of 50 steps within the LACT reconstruction task. As illustrated in \textcolor{subsectioncolor}{Fig.} \ref{omparison of the effectiveness of PWD}, visual comparisons of the ROI regions and corresponding residual maps indicate that PWD achieved more accurate recovery of dental structural details, with improved delineation of edge features and enhanced suppression of artifacts. 
	
	In addition, a quantitative evaluation of reconstruction quality and efficiency across various sampling steps was conducted, as summarized in \textcolor{subsectioncolor}{Table} \ref{tab:comparison}. The results demonstrate that PWD consistently outperformed existing diffusion methods in terms of PSNR and SSIM, while also achieving improved inference speed. These findings validate PWD as a competitive solution that balances reconstruction accuracy and computational efficiency in LACT reconstruction.
\begin{table}[h]
	\centering
	\caption{Quantitative Result of Reconstruction Quality and Reconstruction Efficiency on the Dental Arch Dataset for PWD and Competition Methods with 50 and 200 Sampling Steps.}
	\label{tab:comparison}
	\setlength{\tabcolsep}{5pt} 
	\begin{tabular}{c|ccc|ccc}
		\toprule
		\multirow{2}{*}{Method} & \multicolumn{3}{c|}{$[0^\circ, 90^\circ]$ (200 step)} & \multicolumn{3}{c}{$[0^\circ, 120^\circ]$ (50 step)} \\
		\cmidrule(lr){2-4} \cmidrule(lr){5-7}
		& PSNR & SSIM & Time (s) & PSNR & SSIM & Time (s) \\
		\midrule
		DOLCE & 37.64 & 0.8878 & 35.81 & 36.11 & 0.9068 & 9.44\\
		DDIM  & 35.18 & 0.7957 & 35.62 & 33.06 & 0.8282 & 8.67  \\
		TIFA  & 32.86 & 0.7051 & 65.30 & 32.44 & 0.6910 & 6.05  \\
		DPS   & 34.89 & 0.6854 & 20.62 & 30.26 & 0.6329 & 4.95  \\
		\textbf{Ours} & \textbf{39.65} & \textbf{0.9520} & \textbf{22.49} & \textbf{39.10} & \textbf{0.9464} & \textbf{3.88} \\
		\bottomrule
	\end{tabular}
	\vspace{-10pt}
\end{table}

	\subsection{Ablation Study}
	\subsubsection{Study on Guidance Weight}
\begin{figure}[htb]
		\centering
		\includegraphics[width=0.49\textwidth]{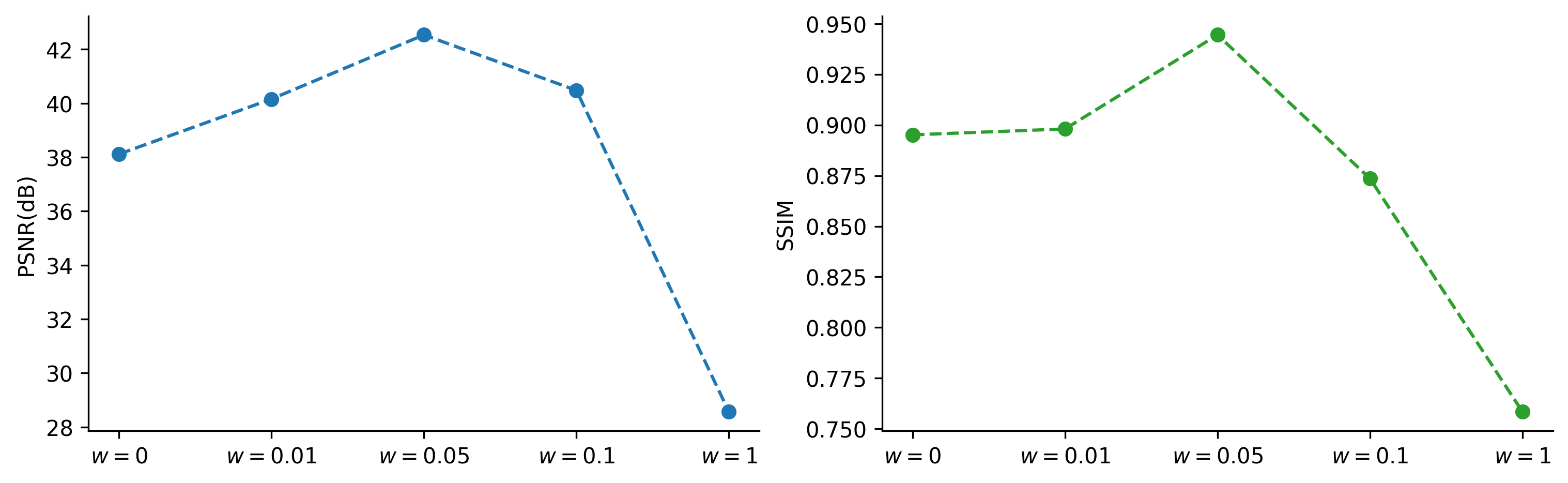} 
		
		\caption{The quantitative comparison of reconstruction performance under 50 steps sampling at  limited angle of 90 degrees. The left and right subfigures respectively show the PSNR and SSIM results corresponding to different guidance weights.}
		\label{guided-Line}
		\vspace{-15pt}
	\end{figure}

	{
		\textcolor{subsectioncolor}{Fig.} \ref{guided-Line} illustrates the impact of different guidance weights $w$ on reconstruction performance under a 50 sampling steps. The left and right subfigures show the trends of PSNR and SSIM, respectively. As observed, the model achieves optimal performance at $w=0.05$, reaching 42.54 dB in PSNR and 0.94 in SSIM. This indicates that moderate structural guidance from LACT images effectively enhances reconstruction quality and improves detail preservation. In contrast, overly large weight $w=1$, leads to excessive dominance of the prior during sampling, suppressing the model’s generative capacity and resulting in a significant performance drop. These findings highlight the importance of balancing guidance strength to ensure both reconstruction stability and image quality.
	}
	
	\vspace{5pt}
	
	\subsubsection{Study on Skip Sampling}
	{
		The impact of sampling step size was systematically evaluated. As shown in \textcolor{subsectioncolor}{Table}~\ref{tab:effect_of_steps}, reconstruction quality with 100 and 50 sampling steps was comparable, while the runtime at 100 sampling steps nearly doubled, indicating diminishing returns in image quality despite higher computational cost. Reducing the steps to 25 caused noticeable quality degradation, especially in edge preservation. At 10 steps, severe artifacts emerged, significantly compromising reconstruction fidelity. Although this configuration tripled inference speed relative to 50 steps, the resulting image quality was inadequate for practical use.
	}
	
	\begin{table}[h]
		\centering
		\caption{Quantitative Results of Using Skip Sampling with Different Numbers of Steps.}
		\label{tab:effect_of_steps}
		\vspace{-5pt}
		\begin{tabular}{c|c|c|c}
			\toprule
			Step number     & PSNR    & SSIM   & Time (s) \\
			\midrule
			100  & 42.69 & 0.9888 & 12.11    \\
			50  & 42.79 & 0.9839 & 3.87     \\
			25  & 40.58 & 0.8995 & 2.61     \\
			10  & 34.94 & 0.8588 & 1.18     \\
			\bottomrule
		\end{tabular}
		\vspace{-5pt}
	\end{table}
	
	\subsubsection{Study on Wavelet Feature Fusion}
\begin{figure}[h]
		\centering
		\includegraphics[width=0.49\textwidth]{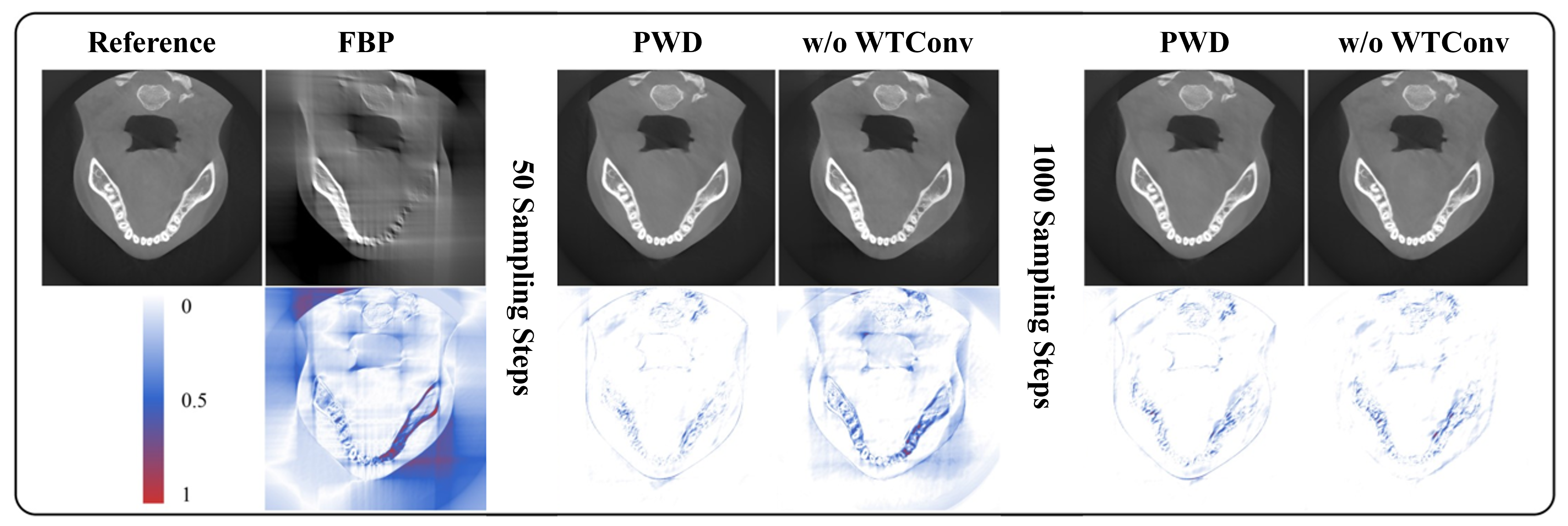} 
		\caption{An ablation study is conducted to evaluate the effectiveness of wavelet-based feature fusion under both 50 steps sampling and full 1000 steps sampling conditions. The display window is [-700, 3400] HU.}
		\label{50-step sampling Yes/No Wavelet comparison}
		\vspace{-10pt}
	\end{figure}

\begin{figure}[htb]
		\centering
		\includegraphics[width=0.49\textwidth]{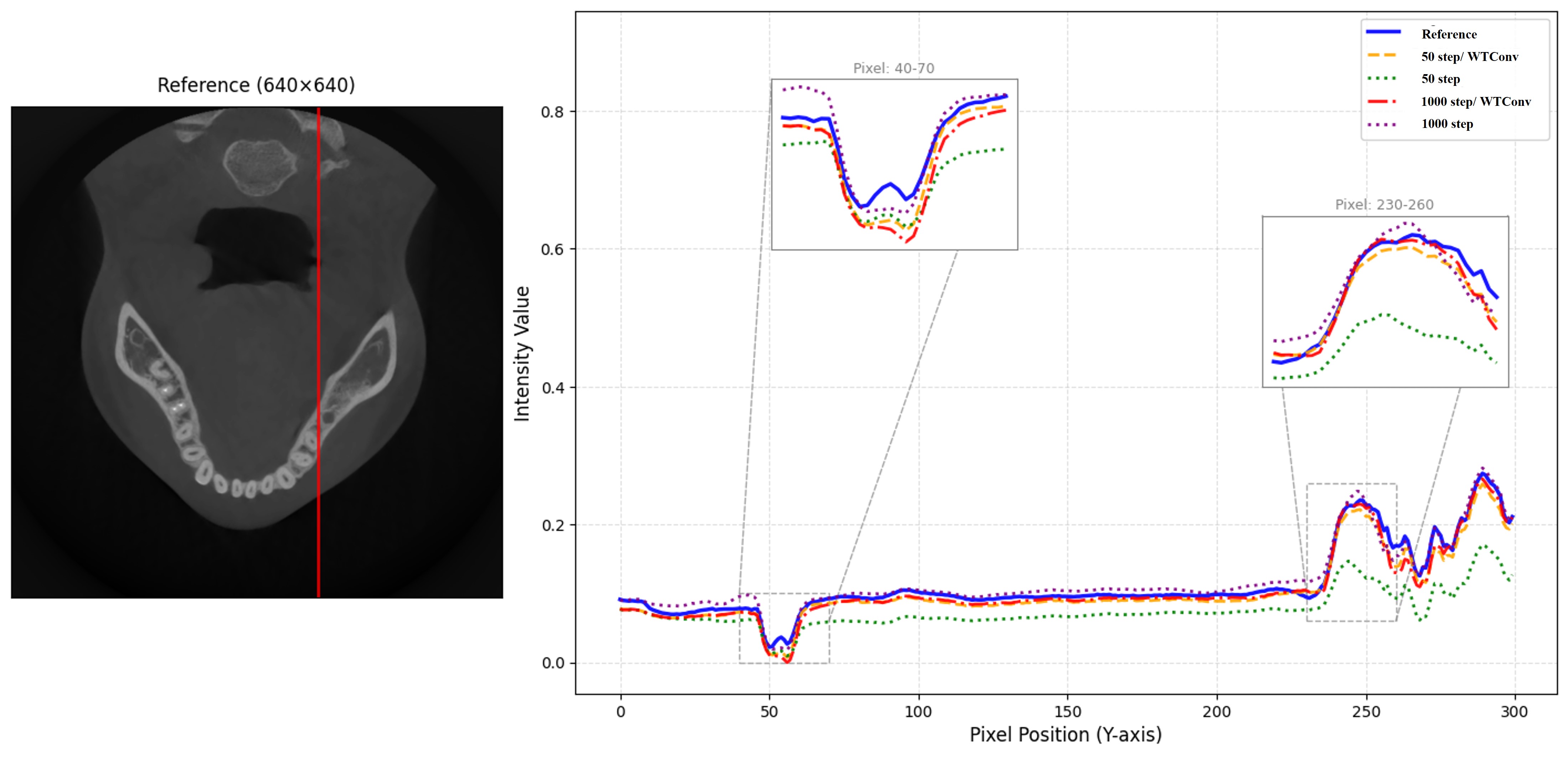} 	
		\caption{Evaluation of the effectiveness of wavelet features through pixel-wise ground truth comparison.}
		\label{Line graph analysis of PWD effects}
		\vspace{-10pt}
	\end{figure}

	{
		
		The comparative experimental results under 50 and 1000 sampling steps conditions, as shown in  \textcolor{subsectioncolor}{Fig.} \ref{50-step sampling Yes/No Wavelet comparison} and \textcolor{subsectioncolor}{Table} \ref{tab:wavelet_ablation}, demonstrate that the introduction of wavelet feature fusion significantly enhances the reconstruction quality of image details under 50 sampling steps. Residual maps further validate that the strategy achieves higher fidelity in the representation of high-frequency areas, such as dental arch edges. In contrast, under 1000 sampling steps, the performance improvement brought by the fusion strategy tends to saturate as the model approaches optimal convergence. The baseline reconstruction method is already capable of recovering most of the image information, thereby reducing the marginal benefit of the fusion strategy. As shown in \textcolor{subsectioncolor}{Fig.} \ref{Line graph analysis of PWD effects}, the red and orange curves representing PWD are closer to the reference, further verifying the effectiveness of wavelet feature fusion in detail recovery.
		
	}

	\begin{table}[h]
		\centering
		\caption{Ablation Study of Feature Fusion in PWD at Different Sampling Steps.}
		\label{tab:wavelet_ablation}
		\begin{tabular}{c|cc|cc}
			\toprule
			\multirow{2}{*}{Step number} & \multicolumn{2}{c|}{PWD} & \multicolumn{2}{c}{PWD (w/o WTConv)} \\
			\cmidrule(lr){2-3} \cmidrule(lr){4-5}
			& PSNR & SSIM & PSNR & SSIM \\
			\midrule
			50  & 37.69 & 0.9451 & 26.95 & 0.6580 \\
			1000 & 38.56 & 0.9552 & 37.26 & 0.8033 \\
			\bottomrule
		\end{tabular}
	\vspace{-10pt}
	\end{table}

	\section{Discussion}
	
	{
		PWD accelerates LACT reconstruction to the second level in this study and achieves high-quality image reconstruction with a limited number of sampling steps. However, certain limitations remain. Since diffusion models rely on a Markov chain structure, further reduction in the number of sampling steps may compromise the stability of the generation process, particularly when using non-Markovian inference schemes such as skip-sampling, which leads to reconstruction failure. Moreover, the current method is primarily designed for two-dimensional slices and does not yet meet the more stringent requirement of millisecond-level reconstruction speed in three-dimensional LACT. Future work will focus on introducing lightweight strategies such as model distillation to further improve reconstruction efficiency in 3D scenarios while preserving image quality.
	
	}
	
	\section{Conclusion}
	{
		This study proposed a fast reconstruction method PWD for LACT. By embedding prior information into both the training and sampling stages of the diffusion model, PWD effectively constrained the solution space, enhanced reconstruction stability, and mitigated the impact of information loss introduced by skipped sampling. In addition, a wavelet-domain feature fusion strategy was incorporated to achieve fine restoration of image structural details. Experimental data were acquired using the Jirox CT system developed by YOFO Medical Technology Co., Ltd, and comprehensive evaluations were conducted. Results demonstrated that the proposed PWD achieved superior reconstruction quality and sampling efficiency in LACT tasks.
	}

	\appendices
	
	\bibliographystyle{IEEEtran}
	\bibliography{sample,reference}

\end{document}